\documentstyle[twoside,12pt]{article}

\def\beq{\begin{equation}} \def\eeq{\end{equation}}
\def\bea{\begin{eqnarray}} \def\eea{\end{eqnarray}}
\let\nn=\nonumber
\def\beann{\begin{eqnarray*}} \def\eeann{\end{eqnarray*}}

   \let\de=\delta
   
  \let\la=\lambda 
  \let\p=\pi  
 \let\ps=\psi
\let\ph=\varphi   
  \let\Th=\Theta

\let\qd=\quad \let\qqd=\qquad \def\qqqd{\qquad\qquad}

\def\tst#1{{\textstyle #1}}
\def\dst#1{{\displaystyle #1}}

\def\0{\over } \def\1{\vec }     \def\2{{1\over2}} \def\4{{1\over4}}
\def\5{\bar }  \def\6{\partial } \def\7#1{{#1}\llap{/}}

\def\<{\langle } \def\>{\rangle }

\let\auf=\uparrow \let\ab=\downarrow

 \def\CS{{\cal S}}
 
\def\i{{\rm i}} \def\rs{{\rm s}}

\renewcommand{\det}{\mbox{det}}

\def\sign{\mbox{sign}}  
\def\mod{\mbox{\,mod\,}}

\def\norm#1{\left| \! \left| \, #1 \, \right| \! \right|}

\begin{document}

\thispagestyle{empty}
\begin{center}
{\Large {\bf The Hubbard chain: Lieb-Wu equations and norm of the
eigenfunctions\\}}
\vspace{7mm}
{\large F.~G\"{o}hmann\footnote[2]
{e-mail: goehmann@insti.physics.sunysb.edu} and
V.~E.~Korepin\footnote[1]{e-mail: korepin@insti.physics.sunysb.edu}}\\
\vspace{5mm}
C. N. Yang Institute for Theoretical Physics,\\ State University of New
York at Stony Brook,\\ Stony Brook, NY 11794-3840, USA\\
\vspace{20mm}

{\large {\bf Abstract}}
\end{center}
\begin{list}{}{\addtolength{\rightmargin}{10mm}
               \addtolength{\topsep}{-5mm}}
\item
We argue that the square of the norm of the Hubbard wave function is
proportional to the determinant of a matrix, which is obtained by
linearization of the Lieb-Wu equations around a solution. This means
that in the vicinity of a solution the Lieb-Wu equations are
non-degenerate, if the corresponding wave function is non-zero. We
further derive an action that generates the Lieb-Wu equations and
express our determinant formula for the square of the norm in terms of
the Hessian determinant of this action.
\\[2ex]
{\it PACS:} 05.30.Fk; 71.10.Pm; 71.27.+a\\
{\it Keywords: Hubbard model; Bethe ansatz; Lieb-Wu equations;
determinant representation}
\end{list}

\clearpage

\subsection*{Introduction}
Interacting many body systems share universal features. The best known
example are the critical exponents which describe the thermodynamic
properties near critical points. Two systems that have the same critical
exponents belong to the same universality class. In one space dimension
there are many systems which are exactly solvable by Bethe ansatz
\cite{Gaudin83,KBIBo,Takahashi99}. Each of these systems represents a
universality class. The Bethe ansatz provides a unique possibility to
obtain exact results for these systems, which are not accessible by
perturbative methods.

In this article we address the problem of the calculation of the norm
of the Bethe ansatz wave functions of the one-dimensional Hubbard model.

In all cases known so far the norms of Bethe ansatz wave functions are
proportional to the determinant of a matrix, which is obtained by
linearization of the Bethe ansatz equations around a solution. This
assures the non-degeneracy of the Bethe ansatz equations in the vicinity
of a solution, if the corresponding wave function does not vanish, and
thus provides an important piece of information about the mathematical
structure of the Bethe ansatz equations.

The first time a determinant formula for the norm of a Bethe ansatz
wave function appeared in the literature was in Gaudin's article
\cite{Gaudin76} on what is nowadays called the Gaudin model. Gaudin's
conjecture was later generalized to the XXX and XXZ spin-$\2$ chains
\cite{MWG81} and to the Bose gas with delta interaction \cite{Gaudin83}.
Due to the complicated nature of the Bethe ansatz wave function a proof
was not available until the advent of the quantum inverse scattering
method, which provided a powerful algebraic reformulation of the Bethe
ansatz.

The first proof of norm formulae for a number of important Bethe ansatz
solvable systems was presented by one of the authors \cite{Korepin82}
in 1982. The proof given in \cite{Korepin82} is rather general. It
relies on the structure of the $R$-matrix and the existence of an
algebraic Bethe ansatz. It applies to the XXX and XXZ spin-$\2$ chains,
to the Bose gas with delta interaction and to a number of other
interesting systems. As a corollary follows the norm formula for the
Gaudin model (for a direct alternative proof of the norm formula
for the Gaudin model see \cite{Sklyanin99}).

In the following years the results of \cite{Korepin82} were generalized
and confirmed by different methods. Reshetikhin \cite{Reshetikhin86}
generalized the proof of \cite{Korepin82} to the case of a $gl_3$
invariant $R$-matrix, i.e.\ to systems with nested Bethe ansatz. Slavnov
proved a formula for the scalar product of a Bethe vector with an
arbitrary vector \cite{Slavnov89}. This formula is valid for models
having the same $R$-matrix as the XXX and XXZ spin-$\2$ chains. It
was recently confirmed by Kitanine, Maillet and Terras who used an
algebraic approach based on Drinfel'd twists \cite{KMT99}. Proofs of
norm formulae which do not rely on the algebraic Bethe ansatz but on the
Knizhnik-Zamolodchikov equation or its quantized version were given by
Reshetikhin and Varchenko \cite{ReVa95} and by Tarasov and Varchenko
\cite{TaVa96}. In the latter paper norm formulae for systems with
$gl_n$ invariant $R$-matrices were obtained.
\subsection*{Bethe ansatz solution of the Hubbard model}
The Hamiltonian of the one-dimensional Hubbard model on a periodic
$L$-site chain may be written as
\beq \label{ham}
     H = - \sum_{j=1}^L \sum_{a = \auf, \ab}
	   (c_{j, a}^+ c_{j+1, a} + c_{j+1, a}^+ c_{j, a})
	   + U \sum_{j=1}^L
	   (n_{j \auf} - \tst{\2})(n_{j \ab} - \tst{\2}) \qd.
\eeq
$c_{j, a}^+$ and $c_{j, a}$ are creation and annihilation operators
of electrons of spin $a$ at site $j$ (electrons in Wannier states),
$n_{j,a} = c_{j, a}^+ c_{j, a}$ is the corresponding particle number
operator, and $U$ is the coupling constant. Periodicity is guaranteed
by setting $c_{L+1, a} = c_{1, a}$.

The eigenvalue problem for the Hubbard Hamiltonian (\ref{ham}) was
solved by Lieb and Wu \cite{LiWu68} by means of the nested Bethe
ansatz (for reviews see \cite{Takahashi99,Sutherland85,DEGKKK99a}).
The Hubbard Hamiltonian conserves the number of electrons $N$
and the number of down spins $M$. The corresponding Schr\"odinger
equation can thus be solved for fixed $N$ and $M$. Since the
Hamiltonian is invariant under a particle-hole transformation and under
reversal of all spins, one may set $2M \le N \le L$ \cite{LiWu68}.

We shall denote the coordinates and spins of the electrons by $x_j$ and
$a_j$, respectively, $x_j = 1, \dots, L$, $a_j = \auf, \ab$. Then the
eigenstates of the Hubbard Hamiltonian (\ref{ham}) may be represented as
\bea \label{state}
     \lefteqn{|N,M\> =} \nn \\ &&
	 \frac{1}{\sqrt{N!}} \sum_{x_1, \dots, x_N = 1}^L\
	      \sum_{a_1, \dots, a_N = \uparrow, \downarrow}
	      \ps (x_1, \dots, x_N; a_1, \dots, a_N) \,
	      c_{x_N a_N}^+ \dots c_{x_1 a_1}^+ |0\>\ .\qd
\eea
$\ps (x_1, \dots, x_N; a_1, \dots, a_N)$ is the Bethe ansatz wave
function. It depends on the relative ordering of the coordinates $x_j$.
Any ordering is related to a permutation $Q$ of the numbers $1, \dots,
N$ through the inequality
\beq \label{sectorq}
     1 \le x_{Q1} \le x_{Q2} \le \dots \le x_{QN} \le L \qd.
\eeq
The set of all permutations of $N$ distinct numbers constitutes the
symmetric group $S_N$. The inequality (\ref{sectorq}) divides the
configuration space of $N$ electrons into $N!$ sectors, which can be
labeled by the permutations~$Q$. The Bethe ansatz wave functions
in the sector $Q$ are given as
\bea
     \lefteqn{\ps (x_1, \dots, x_N; a_1, \dots, a_N) =} \nn \\ && \qqd
	\sum_{P \in S_N} \sign(PQ) \, \ph_P (a_{Q1}, \dots, a_{QN})
	\exp \left( \i \sum_{j=1}^N k_{Pj} x_{Qj} \right) \qd.
	\label{wwf}
\eea
The function $\sign(Q)$ is the sign function (or parity) on the
symmetric group, which is $- 1$ for odd permutations and $+ 1$ for even
permutations.

Explicit expressions for the spin dependent amplitudes $\ph_P (a_{Q1},
\dots, a_{QN})$ can be found in \cite{Woynarovich82a}. They are of the
form of the Bethe ansatz wave functions of an inhomogeneous XXX spin
chain,
\beq \label{wswf}
     \ph_P (a_{Q1}, \dots, a_{QN}) = \sum_{\p \in S_M}
	A(\la_{\p 1}, \dots, \la_{\p M})
	\prod_{l=1}^M F_P (\la_{\p l}; y_l) \qd.
\eeq
Here $F_P (\la; y)$ is defined as
\beq
     F_P (\la; y) = \frac{\i U/2}{\la - \sin k_{Py} + \i U/4}
		    \prod_{j=1}^{y-1} \frac{\la - \sin k_{Pj} - \i U/4}
		    {\la - \sin k_{Pj} + \i U/4} \qd,
\eeq
and the amplitudes $A(\la_1, \dots, \la_M)$ are given by
\beq \label{wwfsa}
     A(\la_1, \dots, \la_M) = \prod_{1 \le m < n \le M}
	\frac{\la_m - \la_n - \i U/2}{\la_m - \la_n} \qd.
\eeq
In the above equations $y_j$ denotes the position of the $j$th down
spin in the sequence $a_{Q1}, \dots, a_{QN}$. The $y$'s are thus
`coordinates of down spins on electrons'. If the number of down spins
in the sequence $a_{Q1}, \dots, a_{QN}$ is different from $M$, the
amplitude $\ph_P (a_{Q1}, \dots, a_{QN})$ vanishes. An expression for
$\ph_P (a_{Q1}, \dots, a_{QN})$ using the terminology of the algebraic
Bethe ansatz can be found in \cite{EKS92a}.

The wave functions (\ref{wwf}) depend on two sets of quantum numbers
$\{k_j \, | \, j = 1, \dots, N \}$ and $\{\la_l \, | \, l = 1,
\dots, M \}$. These quantum numbers may in general be complex. The
$k_j$ and $\la_l$ are called charge momenta and spin rapidities,
respectively. The charge momenta and spin rapidities satisfy the
Lieb-Wu equations (or `periodic boundary conditions')
\bea \label{bak}
     e^{\i k_j L} & = & \prod_{l=1}^M \frac{\la_l - \sin k_j - \i U/4}
                                      {\la_l - \sin k_j + \i U/4} \qd,
		                      \qd j = 1, \dots, N \qd, \\
				      \label{bas}
     \prod_{j=1}^N \frac{\la_l - \sin k_j - \i U/4}
                        {\la_l - \sin k_j + \i U/4} & = &
     \prod_{m=1 \atop m \ne l}^M \frac{\la_l - \la_m - \i U/2}
                        {\la_l - \la_m + \i U/2} \qd,
			\qd l = 1, \dots, M \qd.
\eea

The states (\ref{state}) are joint eigenstates of the Hubbard
Hamiltonian (\ref{ham}) and the momentum operator (cf.\ \cite{GoMu97b})
with eigenvalues
\beq \label{enmom}
     E = - 2 \sum_{j=1}^N \cos k_j + \frac{U}{4}(L - 2N) \qd, \qd
     P = \left( \sum_{j=1}^N k_j \right) \mod \, 2\p \qd.
\eeq

The eigenfunctions (\ref{wwf}) are antisymmetric with respect to
interchange of any two charge momenta $k_j$ and symmetric with respect
to interchange of any two spin rapidities $\la_l$. They are
antisymmetric with respect to simultaneous exchange of spin and space
coordinates of two electrons, and hence respect the Pauli principle.
\subsection*{An action for the Lieb-Wu equations}
Let us rewrite the Lieb-Wu equations (\ref{bak}), (\ref{bas}) in
logarithmic form,
\bea \label{bakl}
     k_j L - \i \sum_{l=1}^M \ln \left(
			     \frac{\i U + 4 (\la_l - \sin k_j)}
			          {\i U - 4 (\la_l - \sin k_j)} \right)
        & = & 2 \p n_j^c \ , \\ \label{basl}
     \i \sum_{j=1}^N \ln \left(
			 \frac{\i U + 4 (\la_l - \sin k_j)}
			      {\i U - 4 (\la_l - \sin k_j)} \right)
     - \i \sum_{m=1}^M \ln \left(
		           \frac{\i U + 2 (\la_l - \la_m)}
			        {\i U - 2 (\la_l - \la_m)} \right)
        & = & 2 \p n_l^s \ . \qd
\eea
Here the logarithm is defined in the cut complex plane, where the cut
is along the real axis from $- \infty$ to zero. $n_j^c$ in equation
(\ref{bakl}) is integer, if $M$ is even and half odd integer, if $M$ is
odd. Similarly, $n_l^s$ in (\ref{basl}) is integer, if $N - M$ is odd,
half odd integer, if $N - M$ is even.

In order to formulate our norm conjecture we shall introduce certain
functions connected to the logarithmic form (\ref{bakl}), (\ref{basl})
of the Lieb-Wu equations. We shall start with the definition
\beq
     \Th_U (x) = \i \int_0^x dy \, \ln \left(
		    \frac{\i U + 4 y}{\i U - 4 y} \right) \qd.
\eeq
In terms of this function the Lieb-Wu equations (\ref{bakl}),
(\ref{basl}) read
\bea
     k_j L - \sum_{l=1}^M \Th_U' (\la_l - \sin k_j) - 2 \p n_j^c & = & 0
	\qd, \\
     \sum_{j=1}^N \Th_U' (\la_l - \sin k_j)
        - \sum_{m=1}^M \Th_{2U}' (\la_l - \la_m)
	- 2 \p n_l^s & = & 0 \qd.
\eea
Here the primes denote derivatives with respect to the argument.
The left hand side of these equations can be easily integrated with
respect to the variables $\sin k_j$ and $\la_l$, yielding the action
\bea
     \CS & = & \sum_{j=1}^N (k_j \sin k_j + \cos k_j) L \nn \\ && \qd
	     + \sum_{j=1}^N \sum_{l=1}^M \Th_U (\la_l - \sin k_j)
	     - \2 \sum_{l,m=1}^M \Th_{2U} (\la_l - \la_m) \nn \\ && \qd
	     - 2 \p \sum_{j=1}^N n_j^c \sin k_j
	     - 2 \p \sum_{l=1}^M n_l^s \la_l \qd.
\eea
Thus, introducing the abbreviation $\rs_j = \sin k_j$, we can write the
Lieb-Wu equations as extremum condition for $\CS$,
\beq \label{sext}
     \frac{\6 \CS}{\6 \rs_j} = 0 \qd, \qd j = 1, \dots, N \qd, \qd
     \frac{\6 \CS}{\6 \la_l} = 0 \qd, \qd l = 1, \dots, M \qd.
\eeq
We shall use the action $\CS$ below in order to formulate our norm
conjecture. A similar action was first introduced by Yang and Yang
in the context of the Bose gas with delta interaction \cite{YaYa69}.

There is an interesting alternative way to write the Lieb-Wu equations.
Let us define
\bea
     \chi_j & = & k_j - \frac{1}{L} \sum_{l=1}^M
		  \Th_U' (\la_l - \sin k_j) - \frac{M \p}{L} \qd, \\
     \ph_l & = & \frac{1}{N} \sum_{j=1}^N \Th_U' (\la_l - \sin k_j)
                 - \frac{1}{N} \sum_{m=1}^M \Th_{2U}' (\la_l - \la_m)
		 - \frac{(N - M + 1)\p}{N} , \qqd
\eea
where $j = 1, \dots, N$ and $l = 1, \dots, M$. In terms of these new
variables the Lieb-Wu equations (\ref{bak}), (\ref{bas}) become
\beq
     e^{\i \chi_j L} = 1 \qd, \qd j = 1, \dots, N \qd, \qd
     e^{\i \ph_l N} = 1 \qd, \qd l = 1, \dots, M \qd.
\eeq
This suggests to interpret the $\chi_j$ and $\ph_l$ as the momenta of
charge and spin degrees of freedom.
\subsection*{The norm formula}
The square of the norm of the wave function (\ref{wwf}) is by definition
\beq
     \norm{\ps}^2 = \sum_{x_1, \dots, x_N = 1}^L \,
	            \sum_{a_1, \dots, a_N = \auf, \ab}
		    |\ps (x_1, \dots, x_N; a_1, \dots, a_N)|^2 \qd.
\eeq
Note that the normalization in (\ref{state}) is such that
\beq
     \<N, M|N, M\> = \norm{\ps}^2 \qd.
\eeq

We suggest the following formula for the square of the norm of the
Hubbard wave function (\ref{wwf}),
\bea
     \norm{\ps}^2 & = & (- 1)^{M'} N! \left( \frac{U}{2} \right)^M
	\prod_{j=1}^N \cos k_j
        \prod_{1 \le j < k \le M} \left(
	       1 + \frac{U^2}{4(\la_j - \la_k)^2} \right) \cdot
	       \nn \\ && \qqd \cdot \,
        \det \left( \begin{array}{cc}
		       \dst{\frac{\6^2 \CS}{\6 \rs^2}} &
		       \dst{\frac{\6^2 \CS}{\6 \rs \6 \la}} \\[2ex]
		       \dst{\frac{\6^2 \CS}{\6 \la \6 \rs}} &
		       \dst{\frac{\6^2 \CS}{\6 \la^2}}
		    \end{array} \right) \qd.
		    \label{norm1}
\eea
Here $M'$ is the number of pairs of complex conjugated $k_j$'s in a
given solution of the Lieb-Wu equations. The determinant on the right
hand side of (\ref{norm1}) is the determinant of an $(N + M) \times
(N + M)$-matrix. This matrix consists of four blocks with matrix
elements
\bea
     \lefteqn{\left( \frac{\6^2 \CS}{\6 \rs^2} \right)_{mn} =
        \frac{\6^2 \CS}{\6 \rs_m \6 \rs_n} \: = \:
	\de_{m,n} \left\{ \frac{L}{\cos k_n} + \sum_{l=1}^M
	   \frac{U/2}{(U/4)^2 + (\la_l - \rs_n)^2} \right\} \qd,} \nn \\
	   && \qqqd \qqqd \qqd m, n = 1, \dots, N \qd, \\
     \lefteqn{\left( \frac{\6^2 \CS}{\6 \la \6 \rs} \right)_{mn} =
     \left( \frac{\6^2 \CS}{\6 \rs \6 \la} \right)_{nm} \: = \:
        \frac{\6^2 \CS}{\6 \la_m \6 \rs_n} \: = \:
	   - \, \frac{U/2}{(U/4)^2 + (\la_m - \rs_n)^2} \qd,} \nn \\
	   && \qqqd \qqqd \qqd m = 1, \dots, M \qd,
	      \qd n = 1, \dots, N \qd, \\
     \lefteqn{\left( \frac{\6^2 \CS}{\6 \la^2} \right)_{mn} =
        \frac{\6^2 \CS}{\6 \la_m \6 \la_n} \: =} \nn \\ && \qqqd
	\de_{m,n} \left\{ \sum_{j=1}^N
	   \frac{U/2}{(U/4)^2 + (\la_n - \rs_j)^2}
	   - \sum_{l=1}^M \frac{U}{(U/2)^2 + (\la_n - \la_l)^2}
	   \right\} \nn \\ && \qqqd
	   + \, \frac{U}{(U/2)^2 + (\la_m - \la_n)^2} \qd, \qd
	   m, n = 1, \dots, M \qd.
\eea
The norm is thus proportional to the Hessian determinant of the action
$\CS$ regarded as a function of the charge momenta $k_j$ and spin
rapidities $\la_l$. Recalling the formulation (\ref{sext}) of the
Lieb-Wu equations we see that a solution is non-degenerate (locally
unique for fixed values of $n_j^c$ and $n_l^s$), if the norm of the
corresponding wave function does not vanish. In other words, all
eigenstates of the Hubbard Hamiltonian (\ref{ham}) correspond to
non-degenerate solutions of the Lieb-Wu equations.

Another interesting form of the norm formula is obtained by expressing
the Hessian determinant in equation (\ref{norm1}) in terms of the
momenta of elementary charge and spin excitations $\chi_j$ and $\ph_l$,
\bea
     \norm{\ps}^2 & = & (- 1)^{M'} N! L^N N^M
			\left( \frac{U}{2} \right)^M
        \prod_{1 \le j < k \le M} \left(
	       1 + \frac{U^2}{4(\la_j - \la_k)^2} \right) \cdot
	       \nn \\ && \qqd
        \cdot \frac{\6 (\chi_1, \dots, \chi_N; \ph_1, \dots, \ph_M)}
	     {\6 (k_1, \dots, k_N; \la_1, \dots, \la_M)} \qd.
	     \label{norm2}
\eea
The norm is proportional to the Jacobian of the transformation from
the set of charge momenta and spin rapidities $k_j$, $\la_l$ to the
set of momenta of charge and spin degrees of freedom $\chi_j$, $\ph_l$.

Let us list our arguments in support of (\ref{norm1}):
\begin{enumerate}
\item
Our experience with other Bethe ansatz solvable models \cite{Korepin82,%
Sklyanin99,Reshetikhin86,TaVa96} shows that norm formulae for Bethe
ansatz wave functions are generically of the form (\ref{norm1}).
\item
It is easy to see that (\ref{norm1}) is valid for $M = 0$ and arbitrary
$N$.
\item
We verified (\ref{norm1}) for $N = 2$, $N = 3$ and $M = 1$. The
calculation is lengthy and involves highly non-trivial cancellations
based on the Lieb-Wu equations.
\item
We verified (\ref{norm1}) for arbitrary $N$ and $M$ in the limit $U
\rightarrow \infty$. This limit requires rescaling of the spin
rapidities, $\lambda_j = \Lambda_j U/2$. The remaining non-trivial
factor in the expression for the norm reduces to the case of the XXX
spin-$\2$ chain, and the result of \cite{Korepin82} can be applied.
\item
We calculated (for arbitrary $N$ and $M$) the leading order term of
the norm in the large $L$ limit. This term is proportional to $L^N$
and fixes the prefactor in (\ref{norm1}).
\end{enumerate}

Finally we would like to point out two applications of the norm
formula (\ref{norm1}). The string hypothesis for the Hubbard model
\cite{Takahashi72} together with the SO(4) symmetry \cite{EKS92a}
predicts the correct number of eigenstates of the Hubbard Hamiltonian
\cite{EKS92b}. This fact is sometimes called combinatorial
completeness. For some Bethe ansatz solvable models there exists an
alternative proof of completeness which avoids the string hypothesis
\cite{TaVa95}. An important ingredient of this proof is a norm formula
similar to (\ref{norm1}).

Over the past two decades a programme was developed for calculating
exactly temperature correlation functions of Bethe ansatz solvable
models \cite{KBIBo}. In the thermodynamic limit the correlation
functions are represented as determinants of Fredholm integral
operators of a certain form. The determinant representation is a
powerful tool for studying the properties of the correlation functions
analytically. It enabled, in particular, the exact calculation of
long and short distance asymptotics for a number of Bethe ansatz
solvable models. The calculation of the norm of the Bethe ansatz
wave function is an important step towards a determinant representation
for correlation functions of the Hubbard model at finite coupling.
For the special case of infinite repulsion a determinant representation 
was recently obtained in \cite{IPA98}.

{\bf Acknowledgement.} This work was supported by the Deutsche
For\-schungsgemeinschaft under grant number Go 825/2-2 (F.G.) and by the
National Science Foundation under grant number PHY-9605226 (V.K.).


\end{document}